\documentclass[11pt]{emulateapj}
\usepackage{epsfig}
\usepackage{amsmath}
\usepackage{multirow}
\begin{document}

\title{Constraining  the emissivity of ultrahigh energy cosmic rays in the distant
universe with the diffuse  gamma-ray emission}

\author{Xiang-Yu Wang\altaffilmark{1,2,4}, Ruo-Yu Liu\altaffilmark{1,4}, Felix Aharonian\altaffilmark{3,2} }
\altaffiltext{1}{Department of Astronomy, Nanjing University,
Nanjing, 210093, China} \altaffiltext{2}{Max-Planck-Institut f\"{u}r
Kernphysik, Saupfercheckweg 1, 69117 Heidelberg, Germany}
\altaffiltext{3}{Dublin Institute for Advanced Studies, 31
Fitzwilliam Place, Dublin 2, Ireland} \altaffiltext{4}{Key
laboratory of Modern Astronomy and Astrophysics (Nanjing
University), Ministry of Education, Nanjing 210093, China}

\begin{abstract}
Ultra-high cosmic rays (UHECRs) with energies  $\ga 10^{19}$ eV
emitted at cosmological distances will be attenuated by cosmic
microwave and infrared background radiation  through photohadronic
processes. Lower energy extra-galactic cosmic rays
($\sim10^{18}-10^{19}$ eV) can only travel a linear distance smaller
than $\sim$Gpc in a Hubble time due to the diffusion if the
extra-galactic magnetic fields are as strong as nano Gauss. These
prevent us from directly observing most of the UHECRs in the
universe, and thus the observed UHECR intensity reflects only the
emissivity in the nearby universe within hundreds of Mpc. However,
UHECRs in the distant universe, through interactions with the cosmic
background photons, produce UHE electrons and gamma-rays that in
turn initiate electromagnetic cascades on cosmic background photons.
This secondary cascade radiation  forms part of the extragalactic
diffuse GeV-TeV gamma-ray radiation and, unlike the original UHECRs,
is observable. Motivated by new measurements of extragalactic
diffuse gamma-ray background radiation by Fermi/LAT, we obtained
upper limit  placed on the UHECR emissivity in the distant universe
by requiring that the cascade radiation they produce not exceed the
observed levels. By comparison with the gamma-ray emissivity of
candidate UHECR sources (such as GRBs and AGNs) at high-redshifts,
we find that the obtained upper limit  for a flat proton spectrum is
$\simeq10^{1.5}$ times larger than the gamma-ray emissivity in GRBs
and $\simeq10$ times smaller than the gamma-ray emissivity in BL Lac
objects.  In the case of iron nuclei composition, the derived upper
limit of UHECR emissivity is a factor of 3-5 times higher. Robust
upper limit on the cosmogenic neutrino flux is further obtained,
which is  marginally reachable by the Icecube detector and the
next-generation detector JEM-EUSO.

\end{abstract}


\section{Introduction}
There is a general consensus that cosmic rays with energy above
$10^{19}$ eV  originate from extragalactic astrophysical sources,
although the sources are unidentified. Some candidates have been
proposed, including active galactic nucleus (AGN) jets (e.g.
Biermann \& Strittmatter 1987; Berezinsky et al. 2006), gamma-ray
bursts (GRBs) (Waxman 1995; Vietri 1995; Wick et al. 2004;  Murase
et al. 2008), and semi-relativistic hypernovae remnants (Wang et al.
2007). Any viable candidates must be able to provide a right amount
of UHECR emissivity  to match the observed flux. Due to the
attenuation by cosmic microwave and infrared background radiation
through photohadronic processes, the energy loss distance of UHECRs
above $10^{19}$ eV is less than several hundreds of Mpc. Although
extragalactic cosmic ray protons  in the lower energy range
$10^{18}\la E\la10^{19}$ eV suffer from much less energy loss, they
could possibly travel a linear distance much smaller than the size
of the universe as well due to the diffusion in the intergalactic
magnetic fields (e.g. Lemoine 2005). For UHECRs whose Larmor radius
$r_L$ is much larger than the coherence scale ($l_c$) of the fields,
which is valid when $E\gg 10^{18} {\rm eV}(B/1{\rm nG})(l_c/1{\rm
Mpc})$, the scattering length is $l_{\rm scatt}\simeq r_L^2/l_c$ and
thus the diffusion coefficient{\footnote{For extragalactic cosmic
ray protons with $E\la10^{18}{\rm eV}(B/1{\rm nG})^{-1}(l_c/{\rm
1Mpc})$, the smaller angle diffusion approximation will not be valid
and the diffusion coefficient in this regime will be different.} is
$D(E)=(1/3)cl_{\rm scatt}=4\times10^{34}(E/10^{18}{\rm
eV})^2(B/1{\rm nG})^{-2}(l_c/{\rm 1Mpc}) {\rm cm^2 s^{-1}}$. In a
Hubble time, these particles travel a linear distance $d\sim(2
H_0^{-1}D)^{1/2}\simeq 60 (E/{\rm 10^{18}eV})(B/1{\rm
nG})^{-1}(l_c/{\rm 1Mpc})^{-1/2}{\rm Mpc}$, where $H_0$ is the
Hubble constant at present. Therefore, the observed UHECR intensity
reflects only the emissivity in the nearby universe within several
hundreds of Mpc. Usually, in the literature this emissivity is
compared with the emissivity in electromagnetic radiation produced
by local candidate sources to check  whether they can provide
sufficient power. The same type of candidate sources are also
present in the distant universe and produce electromagnetic
radiation and UHECRs as well. It would be also useful to compare the
emissivity of UHECRs and emissivity in electromagnetic radiation in
the distant universe, especially when the local sources of the
candidates, such as GRBs or AGN giant flares (Farrar \& Gruzinov
2009), are too infrequent to be detectable in years. While the
emissivity in the electromagnetic radiation can be measured directly
for distant sources, UHECRs produced by them can not be observable
and therefore their emissivity can not be determined directly.
However, UHECRs produce UHE electrons and gamma-rays that in turn
initiate electromagnetic cascades on extragalactic background light
(EBL) and this lower-energy, secondary cascade radiation can be
observable. Thus the energy of UHECRs will be eventually converted
to observable diffuse gamma-rays, as was first noticed by Wdowczyk
et al. (1972). The spectrum of this cascade radiation is rather
insensitive to the spectrum of the original UHECRs (e.g. Strong et
al. 1973; Berezinsky \& Smirnov 1975; Coppi \& Aharonian 1997), and
thus the total level of the cascade background acts as a particle
detector calorimeter, allowing us to measure the total UHECR
emissivity in the universe.

Recently, Kalashev et al. (2009) studied the contribution of the
cascade radiation by UHECRs interacting with EBL photons to the
diffuse extragalactic gamma-ray background (EGBR) measured by EGRET.
Using the new measurement of the EGBR by Fermi/LAT, Berezinsky et
al. (2011) and Ahlers et al. (2010) attempted to constrain the UHECR
source evolution models and made predictions for the flux level of
cosmogenic neutrinos. All these papers assume that the accumulated
UHECR flux from a homogenous population of sources in the whole
universe fit the observed energy spectrum of UHECRs. In our paper,
in order to obtain an independent constraint on the emissivity of
UHECRs in the distant universe, we will relax this assumption since
UHECRs above $10^{19}$ eV are produced dominantly by local sources
within hundreds of Mpc, much smaller than the size of the universe,
whose density could well be enhanced or deficient relative to the
average density. Indeed, there are suggestions that one single
source in the nearby universe, such as Cen A, is the dominated
source producing the observed UHECRs (e.g. Cavallo 1978;   Farrar \&
Piran 2000; Rieger \& Aharonian 2009).

In \S2, we first present the constraints on the cascade emission
imposed by the Fermi/LAT observations of the EGBR. Then in \S3.1, we
present an analytic approach for calculating the energy density of
the cascade emission produced by  UHECRs in the universe. By
comparison with the allowed maximum energy density of the cascade
radiation measured by Fermi/LAT, we obtain upper limits on the UHECR
emissivity at high redshifts in \S3.3. We then discuss the effect of
synchrotron loss of secondary electrons in the presence of
intergalactic magnetic fields on the cascade radiation in \S3.4.
Upper limits on the cosmogenic neutrino flux are further obtained in
\S3.5. We discuss the case of heavy nuclei composition of UHECRs in
\S4. Finally, we give a summary in \S5.

\section{Cascade radiation}
UHECRs interact with EBL photons through photopion process or
Bethe-Heitler process and produce UHE electrons and gamma-rays.
These UHE electrons and gamma-rays interact via Compton and
pair-production process with soft photons in the CMB and radio
background. This will lead to the development of an electromagnetic
cascade, in which the number of electrons and photons increase
quickly. The cascade proceeds until the energy of photons drops
below the pair creation threshold. This process reprocesses
essentially all the energy of UHE electrons and gamma-rays produced
by UHECRs into lower-energy photons below $\sim $ TeV which we can
detect. The cascade low-energy photon spectrum is a universal
spectrum, described by
\begin{equation}
\frac{dn}{d\epsilon_\gamma}\propto\left \{
\begin{array}{ll}
\epsilon_\gamma^{-1.5}\,\, {\rm for} \,\,\epsilon_\gamma<\epsilon_b
\\ \epsilon_\gamma^{-\alpha_\gamma}\,\, {\rm for} \,\,
\epsilon_b<\epsilon_\gamma<\epsilon_{\rm cut}
\end{array}
\right.
\end{equation}
with a steepening at  $\epsilon_\gamma>\epsilon_{\rm cut}$, where
$\epsilon_{\rm cut}$ is the absorption energy of a cascade photon
scattering on EBL, $\epsilon_b=(\epsilon_{\rm cut}/{\rm 1TeV})^2{\rm
GeV}$, and $\alpha_\gamma\simeq 1.8-2$ typically (e.g. Berezinsky \&
Smirnov 1975; Coppi \& Aharonian 1997). The peak of the energy
distribution in this spectrum is at $\epsilon_{\rm cut}$, which is
estimated to be $\sim 100{\rm GeV}$ (Berezinsky et al. 2011).  By
requiring that this theoretical cascade spectrum touches the lower
end of the error bars of the Fermi/LAT data (Abdo et al. 2010),
Berezinsky et al. (2011) obtained an upper limit on the cascade
energy density, i.e.
\begin{equation}
\omega_{\gamma}\la\omega_{\rm cas}^{\rm max}=5.8\times10^{-7} {\rm
eV cm^{-3}}.
\end{equation}
where $\omega_\gamma$ is the cascade energy density produced by any
UHECR source population in the universe.

\section{Constraining the UHECR emissivity with the diffuse gamma-ray emission}
\subsection{Energy loss of UHECRs into the electro-magnetic component}
Since the cascade emission is mainly in the GeV-TeV range and has an
almost universal spectrum, it is numerically much more efficient to
calculate the total energy density $\omega_{\gamma}$ injected into
the cascade and compare this value with the limit $\omega_{\rm
cas}^{\rm max}$ imposed by Fermi/LAT. The cascade energy density is
the accumulation of the electro-magnetic (EM) radiation produced by
the source population over the whole universe. One can calculate it
by summing the contributions by individual sources that are
generated at different cosmological epochs, i.e.
\begin{equation}
\omega_\gamma=\int\int L_p(t)\beta_{0,em}(E_p^s, z(t)) {dt} dE_p^s
\end{equation}
where
\begin{equation}
L_p(t)\equiv \dot{n}(z)E_p^s \frac{dN_p}{dE_p^s},
\end{equation}
is the emissivity of protons per energy decade at energy $E_p^s$ at
some cosmological epoch $t$, $\beta_{0,em}$ is the fraction of the
proton energy deposited into the EM component (photons and pairs)
that we observed at present, $\dot{n}(z)$ is the comoving-frame
number density of protons injected per unit time at time $t=t(z)$,
$E_p^s$ is the proton energy at the source, and
$\frac{dN_p}{dE_p^s}$ is the energy spectrum of protons.

The energy of injected protons evolves with time as
\begin{equation}
-\frac{dE_p}{dt}=E_p H(z)+b(E_p, t)
\end{equation}
where $H(z)$ is the Hubble constant at time $t=t(z)$, and
\begin{equation}
\begin{array} {ll}
b(E_p, t)=\frac{E_p c}{2 \Gamma^2}\int^{\infty}_{\varepsilon_{th}}
d\varepsilon_\gamma \left[\sigma_{\rm BH}(\varepsilon_\gamma)f_{\rm
BH}(\varepsilon_\gamma)
+\sigma_{p\gamma}(\varepsilon_\gamma)f_{p\gamma}(\varepsilon_\gamma)\right]\\
\times \varepsilon_\gamma
 \int^{\infty}_{\varepsilon_\gamma/2\Gamma}d\varepsilon
\frac{n_\gamma(\varepsilon, z)}{\varepsilon^2}
\end{array}
\end{equation}
is the total energy loss rate of protons of energy $E_p$ due to
photopair and photopion interactions with EBL photons at some
cosmological time $t(z)$, $\Gamma$ is the proton Lorentz factor,
$\sigma_{\rm BH}$ and $\sigma_{p\gamma}$ are the cross section for
photopair and photopion production respectively, $f_{\rm BH}$ and
$f_{p\gamma}$ are the fractions of energy loss of protons due to
photopair and photopion interactions in one collision, and
$n_\gamma(\varepsilon, z)$ is the number density of EBL photons of
energy $\varepsilon$ at redshift $z$. We use the cross section in
Chodorowski et al. (1992) for the Bethe-Heitler process and use the
full photopion production cross section from the pion production
threshold up to high energies as described in M\"{u}cke et al.
(2000). The EBL includes CMB photons and infrared-to-optical
background photons (Finke et al. 2010). The number density of
infrared-to-optical background photons at high-redshifts are taken
from the data set
online{\footnote{http://www.phy.ohiou.edu/~finke/EBL/index.html}}.

From the above equations, one can solve $E_p(t)$. Then one can
obtain the energy loss of protons exclusively into the EM component
per unit time at some cosmological time $t(z)$,
\begin{equation}
\begin{array}{ll}
\frac{dE_{p,em}(z)}{dt}=\frac{E_p c}{2
\Gamma^2}\int^{\infty}_{\varepsilon_{th}} d\varepsilon_\gamma
[\sigma_{\rm BH}(\varepsilon_\gamma)f_{\rm BH}(\varepsilon_\gamma)
\\ +  R_{\rm
em}\sigma_{p\gamma}(\varepsilon_\gamma)f_{p\gamma}(\varepsilon_\gamma)]
 \times \varepsilon_\gamma
\int^{\infty}_{\varepsilon_\gamma/2\Gamma}d\varepsilon
\frac{n_\gamma(\varepsilon, z)}{\varepsilon^2}
\end{array}
\end{equation}
where $R_{em}\simeq 0.6$ is the fraction of proton energy that goes
into the EM component in the photopion channel (the other
$\simeq0.4$ goes into the neutrino production channel) (Engel et al.
2001).

Then the energy  lost into the EM component by a proton of energy
$E_p^s$ (at the source) during the whole period from the injection
time $t(z)$ to the present time is
\begin{equation}
\beta_{em}(E_p^s, t)E_p^s\equiv \int_0^{t(z)}
\frac{dE_{p,em}(z)}{dt} dt= \int_0^z \frac{dE_{p,em}(z)}{dt}
\frac{dt}{dz}dz.
\end{equation}
The corresponding energy that remains  at the present epoch after
taking into account the redshift energy loss is
\begin{equation}
\beta_{0,em}(E_p^s, t)E_p^s\equiv \int_0^{t(z)}
\frac{1}{1+z}\frac{dE_{p,em}(z)}{dt} dt.
\end{equation}
The values of $\beta_{em}(E_p^s, t)$ as a function of the energy of
protons generated at different redshifts  are shown in Fig.1. For a
source at $z\ga 0.5$, more than a half of the proton energy is lost
into the EM cascade for protons with energy above $10^{18.5}$ eV.

\subsection{The UHECR source density evolution}
The evolution of the UHECR emissivity per energy decade with
redshift can be parameterized as
\begin{equation}
L_p(z)=L_p(z=1)\frac{S(z)}{S(z=1)},
\end{equation}
where $L_p(z=1)\equiv \dot n(z=1) E_p^s \frac{dN_p}{dE_p^s}$,
$\dot{n}(z=1)$ is the comoving-frame number density of protons
injected per unit time at  $z=1$ and $S(z)$ is the source density at
redshift $z$. We take three cases for $S(z)$ in the following
calculation, i.e. $S(z)$ follows the star formation history (SFR),
gamma-ray burst (GRB) rate and active galactic nuclei (AGN) rate in
the universe respectively.

We take the form of SFR from Y\"{u}skel et al. (2008)
\begin{equation}
S_{\rm SFH}(z)\propto \left \{
\begin{array}{lll}
(1+z)^{3.4}, \,\,\,\, z<1 \\
(1+z)^{-0.3}, \,\,\,1<z<4\\
(1+z)^{-3.5}. \,\,\,\,z>4
\end{array} \right .
\end{equation}

Recent analysis of the GRB redshift distribution as detected by {\it
Swift} reveals that the GRB rate is enhanced at high redshift
relative to SFR (Le \& Dermer 2007). This may arise from some
mechanisms (Kistler et al. 2008), such as a GRB preference for
low-metallicity environments (Stanek et al. 2006; Langer \& Norman
2006). Following Yuksel \& Kistler (2007), we assume  $S_{\rm
GRB}(z)\propto (1+z)^{1.4}S_{\rm SFH}$, which gives
\begin{equation}
S_{\rm GRB}(z)\propto \left \{
\begin{array}{lll}
(1+z)^{4.8}, \,\,\, z<1 \\
(1+z)^{1.1}, \,\,\,1<z<4\\
(1+z)^{-2.1}. \,\,\,z>4
\end{array} \right .
\end{equation}
As shown by Yuksel \& Kistler (2007), this source density evolution
function is consistent with the evolution function obtained in Le \&
Dermer (2007).

AGNs may have a similarly strong evolution with redshift, as found
in Hasinger et al. (2005) for different luminosity AGNs. Following
Hasinger et al. (2005) and Ahlers et al. (2009), we take the form of
\begin{equation}
S_{\rm AGN}(z)\propto \left \{
\begin{array}{lll}
(1+z)^{5.0},  \,\,\,z<1.7 \\
{\rm constant}, \,\,\,1.7<z<2.7\\
10^{(2.7-z)}. \,\,\,\,z>2.7
\end{array} \right .
\end{equation}

\subsection{Constraints on the emissivity of UHECRs in the distant
universe} Once  we know $\beta_{0,em}$ and the source density
evolution function $S(z)$, we can calculate the cascade energy
density $\omega_\gamma$, i.e.
\begin{equation}
\omega_\gamma=\int_{0}^{z_{max}}\int_{E_{p,\rm min}}^{E_{p,\rm max}}
\beta_{0,em}(E_p^s, z)L_p(z=1)\frac{S(z)}{S(z=1)} \frac{dt}{dz}
dE_p^s  dz,
\end{equation}
where $E_{p,\rm min}$ and $E_{p,\rm max}$ are the minimum and
maximum energy of extragalactic cosmic rays,  $z_{max}$ is the
maximum redshift of  extragalactic cosmic ray sources, and
${dz}/{dt}=H_0(1+z)[\Omega_M(1+z)^3+\Omega_\Lambda]^{1/2}$ (with
$\Omega=0.3$, $\Omega_\Lambda=0.7$, and $H_0=70 {\rm km s^{-1}
Mpc^{-1}}$). We assume a power-law spectrum for extragalactic cosmic
rays between $E_{p,\rm min}$ and $E_{p,\rm max}$, i.e.
${dN_p}/{dE_p^s}\propto (E_p^s)^{-\gamma_g}$. As $\beta_{em}$
becomes negligible at $E_p<10^{17}{\rm eV}$,  $E_{p,min}$ is taken
to be $10^{17}{\rm eV}$ in  the following calculations, unless
otherwise specified.

Requiring $\omega_{\gamma}\la\omega_{\rm cas}^{\rm max}$, one can
obtain the upper limit on the UHECR emissivity  at  $z=1$ for
different source density evolution scenarios. We define the total
UHECR emissivity as the integral  of $L_p(z=1)$ over the energy
range from $E_{p,min}$  to $E_{p,max}$, i.e.
\begin{equation}
P(z=1)\equiv\int_{E_{p,min}}^{E_{p,\rm max}} L_p(z=1)dE_p^s.
\end{equation}

The upper limits of $P(z=1)$ for different spectral index $\gamma_g$
are shown in Fig.2 (the upper panel) with a fixed maximum proton
energy $E_{p,max}=10^{21}{\rm eV}$ (but different minimum energies)
for different source density evolution scenarios. The values are in
the range of a few $10^{45}$ to a few $10^{46} {\rm erg Mpc^{-3}
yr^{-1}}$. The upper limit for the SFH case is the highest because
of the slowest redshift evolution and smaller $\beta_{0,em}$ at
lower redshifts. As a comparison, we also show the gamma-ray
emissivity of GRBs and BL Lac objects at redshift $z=1$. Though the
local gamma-ray emissivity of GRBs inferred by different groups are
different by about one order of magnitude (Guetta et al. 2005; Le \&
Dermer 2007; Wanderman \& Piran 2010), the inferred emissivity in
GRBs at redshift $z=1$ has a much smaller uncertainty, being
$1.1\times10^{44}(\Delta t/10 {\rm s}){\rm erg Mpc^{-3} yr^{-1}}$ in
Guetta et al. (2005), $1.8\times10^{44}(\Delta t/10 {\rm s}){\rm erg
Mpc^{-3} yr^{-1}}$ in Wanderman \& Piran (2010), and
$2.5\times10^{44}(\Delta t/10 {\rm s}){\rm erg Mpc^{-3} yr^{-1}}$ in
Le \& Dermer (2007), where $\Delta t$ is the mean duration of long
GRB in the explosion frame. By comparison with the upper limits on
the UHECR emissivity, we find that the ratio between the two
emissivities is $P_{\rm CR}(10^{17}-10^{21}{\rm eV})/P_{\gamma}\la
10^{1.5}$ for a flat cosmic ray spectrum $\gamma_g\simeq2.0$.

Dermer \& Razzaque (2010) recently obtained the gamma-ray emissivity
of different classes of AGNs. The two bright classes at high
redshift are BL Lac objects and flat spectrum radio quasars (FSRQs).
The gamma-ray emissivity in BL Lac objects and FSRQs at $z=1$ are,
respectively,  $6\times10^{46}{\rm erg Mpc^{-3} yr^{-1}}$ and
$1.5\times10^{46}{\rm erg Mpc^{-3}}$. If these sources produce
UHECRs, we have $P_{\rm CR}(10^{17}-10^{21}{\rm eV})/P_{\gamma}\la
0.1$ for $\gamma_g\simeq2.0$. This suggests that either a small
fraction of the BL Lac objects in the distant universe are capable
of accelerating protons to ultra-high energies or the
baryon-electron ratio in these sources is significantly smaller than
one.

Though there are few astrophysical sources that can accelerate
protons to energies beyond $10^{20}{\rm eV}$, there should be more
sources in the universe that can accelerate protons to a lower
$E_{p,max}$. For example, accretion shocks in clusters of galaxies
may be able to accelerate protons to $10^{18}-10^{19}{\rm eV}$ (e.g.
Inoue et al. 2005). These ultrahigh energy protons produce diffuse
gamma-ray emission as well. Therefore, the observed extragalactic
diffuse gamma-ray background also constrains the UHECR emissivity
produced by these lower $E_{p,max}$ accelerators. The upper limits
of UHECR emissivity for different $E_{p,max}$ are shown in Fig.3. It
shows that for lower $E_{p,max}$ accelerators, the   UHECR
emissivity could be higher, but it must still  be lower than
$\sim10^{47}{\rm erg Mpc^{-3} yr^{-1}}$ at $z=1$.

%In the above, we assumed a fixed minimum energy at
%$E_{p,min}=10^{17}{\rm eV}$ for the UHECR spectrum. In order to see
%how the results depend on the the assumption made on $E_{p,min}$, we
%show the upper limits on the UHECR emissivity for
%$E_{p,min}=10^{18}{\rm eV}$ and $E_{p,min}=10^{19}{\rm eV}$ in
%Fig.3.

\subsection{The effect of intergalactic magnetic field on the cascade radiation}

When the intergalactic magnetic fields are strong, the synchrotron
cooling of the secondary electrons produced by UHECRs propagating in
the intergalactic space could dominate over the inverse-Compton
cooling and as a result, the cascade energy is reduced to some
extent compared to the case without the magnetic fields.
Unfortunately, very little is known about the origin, spatial
configuration and amplitude of the intergalactic magnetic fields.
Only upper limit on the intergalactic magnetic fields is obtained
from Faraday rotation measurements, i.e. $B\la10^{-8}{\rm G}
(l_c/{\rm 1Mpc})^{-1/2}$, where $l_c$ is the coherence length scale
of the fields (e.g. Ryu et al. 1998). Following Gabici \& Aharonian
(2005) and Kotera et al. (2011), the effective inverse Compton
cooling time on the CMB and radio backgrounds can be written as
$t_{e\gamma}\simeq 5\times10^{14} {\rm s}(E_e/10^{18}{\rm
eV})^{\alpha_{IC}}$, with ${\alpha_{IC}}$=1 if the electron energy
$E_e\la10^{18}{\rm eV}$ and ${\alpha_{IC}}$=0.25 if $10^{18}{\rm
eV}\la E_e\la10^{20}{\rm eV}$. Above $10^{20}{\rm eV}$,
$t_{e\gamma}\simeq 1.6\times10^{15} {\rm s}(E_e/10^{20}{\rm eV})$.
In comparison, the synchrotron cooling time is
$t_{eB}\simeq4\times10^{14}{\rm s}(B/{\rm 1nG})^{-2}(E_e/10^{18}{\rm
eV})^{-1}$. The opposite scalings of $t_{e\gamma}$ and $t_{eB}$ with
electron energy imply the existence of a cross-over energy
$E_{e,cr}$, above which electrons cool mainly via synchrotron
radiation instead of undergoing an inverse Compton cascade, i.e.
$E_{e,cr}=10^{18}{\rm eV} B_{\rm nG}^{-1}$ for $B\ga 1 {\rm nG}$,
$E_{e,cr}=10^{18}{\rm eV} B_{\rm nG}^{-1.6}$ for $1 {\rm nG}\ga B\ga
0.1 {\rm nG}$, and $E_{e,cr}=5\times10^{20}{\rm eV} B_{0.01 \rm
nG}^{-1}$ for $B\la 0.1 {\rm nG}$ (Kotera et al. 2011). Since the
energy of secondary electrons produced by UHE protons is a fraction
of 1/20 or $10^{-3}$ of the parent proton energy in the photopion or
photopair process, electrons with energies above
$E_{e,cr}\ga10^{18}{\rm eV}$ can be produced only by the photopion
process of UHE protons. In Fig.4, we show the fraction  energy loss
of UHE protons  into the cascade EM component (i.e. excluding the
synchrotron radiation) in the presence of the intergalactic magnetic
fields (for three cases of $B=$ 0.01, 0.1 and 1 nG ). It shows that
this fraction drops significantly as the magnetic fields increase
for protons above the photopion threshold energy. We further
calculate the cascade energy density $\omega^B_\gamma$  in the
presence of intergalactic magnetic fields. In Fig.5, we show the
ratio of the cascade energy densities in the presence of a magnetic
field ($\omega^B_\gamma$) and in its absence ($\omega_\gamma$) as a
function of the maximum proton energy for different UHECR source
populations. When $E_{p,max}$ is large, up to $\sim45\%$ of the
cascade energy is lost into the synchrotron radiation.

The characteristic energy of the synchrotron radiation is
$E_{\gamma,syn}\simeq 6.8 {\rm GeV} B_{nG}(E_e/10^{19}{\rm eV})^2$.
So when the magnetic fields are stronger, e.g. $1 {\rm nG}<B<10{\rm
nG}$, the synchrotron peak falls at 10-100 GeV. Thus the synchrotron
component in fact also contributes to EGRB, although its spectrum is
different from that of the cascade radiation.  Therefore we conclude
that the presence of intergalactic magnetic fields affects the upper
limits of the UHECR emissivity by a factor $\la2$.

\subsection{The  upper bound on the cosmogenic neutrino flux}
Once we know the upper bound on the UHECR emissivity, we can get the
upper bound on the cosmogenic neutrino flux. As the energy loss
distance for protons of energy above the photopion production
threshold in interactions with CMB photons  is relatively short
($\la 300$ Mpc), it is reasonable to assume that cosmic rays at
these energies lose all of their energy locally. Following Engel et
al. (2001) and Yuksel \& Kistler (2007), we use the approximation
that the fraction of the original proton energy that is lost to
neutrinos can be parametrized with a gradual step function
$\psi(E_p)=0.45/(1+(E_t/E_p)^2)$, where 0.45 is the asymptotic
fraction of injected cosmic-ray energy transferred to neutrinos
above $10^{21}$ eV and $E_t\sim 2\times10^{20}{\rm eV}$. Thus, the
total $\nu_\mu+\bar\nu_\mu$ energy flux at Earth can be written as a
sum of the contributions by individual UHECR sources that are
generated at different cosmological epochs, i.e.
\begin{equation}
E_\nu J_\nu=\frac{c}{4\pi}\int_0^{z_{max}}\frac{1}{1+z}
\psi(E_p^s(1+z)) L_p(t)\frac{dE_p^s}{dE_\nu} \frac{dt}{dz}dz,
\end{equation}
where $E_p^s =20(1+z)E_\nu$ (approximating each daughter neutrino
receiving about 1/20 of the injected proton energy).

The results are shown in Fig.6 for $E_{p,min}=10^{17}$ eV,
$E_{max}=10^{21}$ eV and a flat proton spectrum with $\gamma_g=2$.
In contrast to Berezinsky et al. (2011) and Ahlers et al. (2010), we
relax the requirement that the accumulated UHECR flux from a
homogenous population of sources in the whole universe fit the
observed energy spectrum of UHECRs since UHECRs above $10^{19}$ eV
are produced dominantly by local sources within hundreds of Mpc and
the UHECR emissivity in the distant universe could be irrelevant. As
a result, the upper bound on cosmogenic neutrino flux shown in Fig.6
represents a true upper bound that is independent of the unknown
density distribution of nearby sources that contribute to the
observed UHECR flux. This upper bound is below the sensitivity of
Auger and is marginally reachable by Icecube and the next-generation
detector JEM-EUSO.

The commonly-used Waxman-Bahcall bound (Waxman \& Bahcall 1999) for
high-energy neutrinos  also assumes that UHECR emissivity in the
distant universe is connected with the  flux of observed UHECRs that
are produced by local sources. Without any independent constraints,
the UHECR emissivity in the distant universe could in principal be
arbitrarily high. However, with the upper limit of the cascade
radiation, the constraints on the UHECR emissivity in the distant
universe becomes possible now and hence the neutrino upper bound
becomes solid.

\section {The iron nuclei case}
The composition of UHECRs remains disputed. Although HiRes
observations favor proton composition (Abbasi et al. 2010), recent
observations by Pierre Auger Observatory (PAO) show a transition in
the maximum shower elongations ${\rm <X_{max}>}$ and in their
fluctuations ${\rm RMS(X_{max})}$ between 5EeV and 10EeV (Abraham et
al. 2010), which are interpreted as reflecting a transition in the
composition of UHECR in this energy range from protons to heavier
mass nuclei. However, one should be cautious that this claim depends
on the poorly-understood hadronic interaction models at such high
energies. In this section, we study the heavy nuclei composition
case. For simplicity,  a pure iron composition above $10^{19}$ eV is
assumed in our calculation and the maximum energy of iron nuclei is
fixed at $10^{21}$ eV. As the corresponding maximum energy of one
nucleon is only $1.8\times10^{19}$ eV, which is below the threshold
energy for photopion production of protons interacting with CMB
photons even at high redshifts, we can neglect the photopion energy
loss for these UHE iron nuclei.

For an Fe nucleus generated at cosmological time $t(z)$, it suffers
from both energy loss due to Bethe-Heitler process and nucleon loss
due to photo-disintegration during the propagation in the
intergalactic space. Let's denote $\gamma_N(t)$ as the Lorentz
factor  of the nucleus and $A(t)$ as the mass number of the nucleus.
As the Lorentz factor  of the nucleus is conserved during
photo-disintegration, the energy loss is due to Bethe-Heitler
cooling and adiabatic expansion of the universe. The energy loss due
to de-excitation of nuclei following photo-disintegration
interactions is found to be always less efficient than the
Bethe-Heitler energy loss (Aharonian \& Taylor 2010), so it can be
safely neglected. So the the evolution of $\gamma_N(t)$ with time is
given by
\begin{equation}
\frac{d\gamma_N(t)}{dt}=\gamma_N(t)H(z)+ \dot{\gamma}_{N,BH}(t,A),
\end{equation}
where
\begin{equation}
\begin{array}{ll}
\dot{\gamma}_{N,BH}(t,A)=\frac{Z^2}{A} \dot{\gamma}_{p,BH}(t)
\\ =\frac{Z^2}{A}\frac{c}{2 \gamma_N}\int^{\infty}_{\varepsilon_{th}}
d\varepsilon_\gamma
\sigma_{BH}(\varepsilon_\gamma)f(\varepsilon_\gamma)\varepsilon_\gamma
\int^{\infty}_{\varepsilon_\gamma/2\Gamma}d\varepsilon
\frac{n_\gamma(\varepsilon)}{\varepsilon^2}
\end{array}
\end{equation}
is the Bethe-Heitler energy loss rate for nucleus of charge $Z$ and
mass number $A$, $\dot{\gamma}_{p,BH}(t)$ is the Bethe-Heitler
energy loss rate for protons of the same Lorentz factor $\gamma_N$.
The photo-disintegration results in nucleon loss of nuclei, so the
mass number evolves with time as
\begin{equation}
\frac{dA(t)}{dt}=R_A(t,\gamma_N),
\end{equation}
where
\begin{equation}
%\begin{array}{ll}
R_A(t,\gamma_N)=\frac{c}{2\gamma_N^2}\int_{\varepsilon_{{\rm
th}}}^\infty d\varepsilon\sigma_{{ A}} (\varepsilon)\varepsilon
\int_{\varepsilon/2\gamma_N}^{\infty}dx x^{-2}n_\gamma(x)
%\end{array}
\end{equation}
is the total photo-disintegration rate for nucleus with Lorentz
factor $\gamma_N$ and mass number $A$, $\sigma_{{A}}(\epsilon)$ is
the total photodisintegration cross section, and $\varepsilon_{{\rm
th}}$ is the threshold energy of the photon in the nucleus rest
frame.

It is found that the single nucleon loss is the dominant channel for
photo-disintegration of heavy nuclei (Puget et al. 1976), so as a
good approximation, we here only consider  single nucleon loss
channel in the following calculation. This approximation results in
an error less than $30\%$ for nucleus with Lorentz factor smaller
than $2\times10^{10}$ (corresponding to energy $\la10^{21}{\rm eV}$
for an iron nucleus)(Puget et al. 1976). The cross-sections for
photodisintegration in the energy range $\varepsilon_{th} <
\varepsilon \lesssim 30$~MeV with single nucleon loss is dominated
by the giant dipole resonance (GDR), which can be approximately
described by a Lorentzian form (Puget et al. 1976; Anchordoqui et
al. 2007) as
\begin{equation}
\sigma_A(\varepsilon)=\frac{\sigma_{0,A}
{\varepsilon}^2\Delta_{GDR}^2}{({\varepsilon}_0^2-{\varepsilon}^2)^2
+{\varepsilon}^2\Delta_{GDR}^2},
\end{equation}
where $\Delta_{GDR}$ and $\sigma_{0,A}$ are the width and maximum
value of the cross section, $\varepsilon_0$ is the energy at which
the cross section peaks.  Fitted numerical values are
$\sigma_{0,A}=1.45A\times10^{-27}~{\rm cm^2}$, $\Delta_{GDR}=8~{\rm
MeV}$, and $\varepsilon_0=42.65 A^{-0.21}$($0.925A^{2.433}$)~MeV for
$A>4$ ($A<4$) (Karakula \& Tkaczyk 1993).

Combining Eq.(18) and Eq.(20), one can obtain $\gamma_N(t)$ and
$A(t)$.  Denote $t_{A=1}$ as the time when the mass number of a
parent nucleus drops to $A=1$. Then, one can calculate the  energy
lost into the EM component by an Fe nucleus with an initial energy
$E_N^s$ during the whole period from the the injection time $t(z)$
to the present time
\begin{equation}
\begin{array}{ll}
\beta_{em}^N(E_N^s, t)E_N^s\equiv \int_{t_{A=1}}^{t(z)}
[\dot{\gamma}_{N,BH}(t,A) A(t)+ (56-A) \\
\times\dot{\gamma}_{p,BH}(t)] m_p c^2dt+\int_0^{t_{A=1}}
[56\dot{\gamma}_{p,BH}(t)]m_p c^2dt,
\end{array}
\end{equation}
where the second term on the right side is the Bethe-Heitler energy
loss for those secondary nucleons that have been already
disintegrated from the parent nucleus. Note that, for the sake of
analytic calculations, we have assumed that all the secondary
nucleons that have been disintegrated from the parent nucleus have
an averaged $\gamma_p$ at some time $t$, which is a good
approximation since the photo-disintegration process of nuclei in
the majority of the relevant energy range is faster than the
Bethe-Heitler cooling  (Puget et al. 1976; Stecker \& Salamon 1999;
Ave et al. 2005)\footnote{In the small energy range where the
Bethe-Heitler cooling is more efficient (e.g. Ave et al. 2005), the
energy loss of parent nucleus  into the EM cascade dominate over the
energy loss by the disintegrated nucleons since the Bethe-Heitler
cooling efficiency scales as $Z^2/A$, therefore Eq.(22) remains a
good approximation.}. In Fig.7, we show $\beta_{em}^N$ as a function
of the energy of iron nucleus generated at different redshifts. Note
that the bumps in these curves correspond to the energies of those
iron nuclei that interact with  CMB photons at the GDR peak. The
rise of $\beta_{em}^N$ at higher energies is due to the
contributions by the nucleons that have been disintegrated from the
parent nucleus. For an iron nucleus generated at $z\ga1$ with an
energy $\ga5\times10^{19}$ eV, more than a half of the nucleus
energy  is lost into the EM cascade radiation. Compared with the
proton case, the energy loss fraction is smaller at energies $E_{\rm
Fe}\la10^{20}$ eV.

Similar to the proton case, we can calculate the corresponding
energy that remains at the present epoch, $\beta_{0,em}^NE_N^s$, by
taking into account the redshift energy loss. Finally, we obtain the
cascade energy density produced by the whole UHECR source population
in the universe
\begin{equation}
\omega_{\gamma,N}=\int_{0}^{z_{max}}\int_{E_{N,\rm min}}^{E_{N,\rm
max}} \beta_{0,em}^N(E_N^s, z(t))L_N(z=1)  \frac{S(z)}{S(z=1)}
\frac{dt}{dz} dE_N^s dz,
\end{equation}
where $L_N(z=1)$ is the emissivity of UHE iron nuclei per energy
decade at $z=1$. Requiring $\omega_{\gamma, N}\la\omega_{\rm
cas}^{\rm max}$, we  obtain the upper limits on the UHECR emissivity
at redshift $z=1$ (i.e. $P(z=1)$)  for different source evolution
scenarios, which are shown in Fig.8. Note that for the iron
composition case, $P(z=1)$ is defined as the integral of $L_N(z=1)$
over the energy range from $10^{19}$ eV to $10^{21}$ eV. These upper
limits are a factor of 3-5 times higher than that in the proton
composition case due to smaller energy loss fractions
$\beta^N_{0,em}$ at energies $E_{\rm Fe}\la10^{20}$ eV.

\section{Summary and Discussions}
The observed UHECR intensity above $10^{19}$ eV reflects only the
UHECR emissivity in the nearby universe due to that these extremely
high-energy particles lose energy quickly while propagating in the
universe. Thus, there is no direct information about the UHECR
emissivity in the distant universe. In this paper, we suggest that
the cascade gamma-ray radiation initiated by the secondary particles
produced by UHECRs provides a useful probe of the UHECR emissivity
in the distant universe, since this cascade gamma-ray radiation can
be observed directly. Using the new Fermi/LAT measurement of the
EGBR, we obtained upper limits on the UHECR emissivity in the
distant universe through an analytic treatment. Both proton
composition and pure iron composition cases are studied. These
limits are then compared with the  gamma-ray emissivity of candidate
UHECR sources, i.e. GRBs and AGNs, at high redshifts. We find that,
for a flat proton spectrum, the upper limit of the UHECR emissivity
is a few tens times larger than the observed gamma-ray emissivity in
GRBs, while it is about one order of magnitude  smaller than the
gamma-ray emissivity in BL Lac objects. We also find that the
presence of intergalactic magnetic fields has insignificant effect
on the derived upper limit. Furthermore, an upper limit on the
cosmogenic neutrino flux is obtained from the upper limit of UHECR
emissivity. In contrast to Berezinsky et al. (2011) and Ahlers et
al. (2011), we relax the assumption that the accumulated UHECR flux
from a homogenous population of sources in the whole universe fit
the observed UHECR energy spectrum, since UHECRs above $10^{19}$ eV
originate from local sources within hundreds of Mpc and the UHECR
emissivity in the distant universe could be irrelevant. Therefore
our upper limit on the cosmogenic neutrino flux  represents a true
upper bound that is independent of  the unknown density distribution
of the nearby sources that contribute to the observed UHECR flux.
This upper bound is below the sensitivity of Auger and is only
marginally reachable by Icecube and the future detector JEM-EUSO.

Recently, Neronov \& Semikoz (2011) obtained the flux of the EGBR
using a different method from that used in Abdo et al. (2010), which
is a factor of $\simeq 2$ below the flux obtained  in Abdo et al.
(2010) in the 100-200 GeV range. If this flux is true, the upper
limits on the UHECR emissivity as well as the upper bounds on the
cosmogenic neutrinos will go down by the same  factor of $\simeq2$.

\acknowledgments  We thank the referee for the helpful comments and
suggestions. This work is supported by the NSFC under grants
10973008 and 11033002, the 973 program under grant 2009CB824800, the
Program for New Century Excellent Talents in University, the Qing
Lan Project and the Fok Ying Tung Education Foundation.

\newpage

\begin{figure}
\centering \epsfig{figure=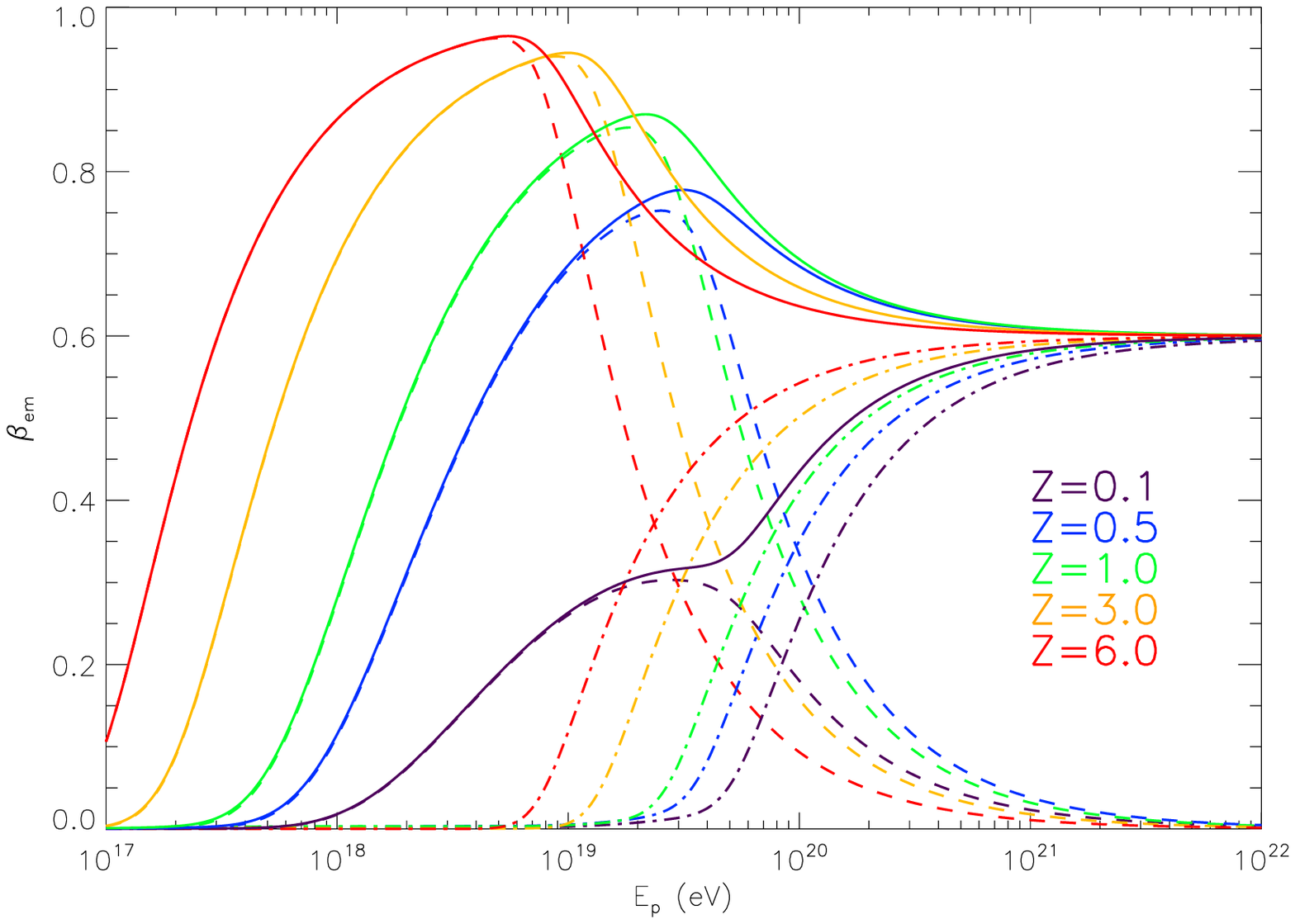,width=12cm} \caption{The fraction
energy loss of UHE protons into the electro-magnetic component for
protons generated at different redshifts. The dashed lines and dash
dotted lines represent the fractions of energy loss through the
photopair channel and photopion channel respectively, while the
solid lines represent the sum of them. \label{fig1}}
\end{figure}

\begin{figure}
\centering \epsfig{figure=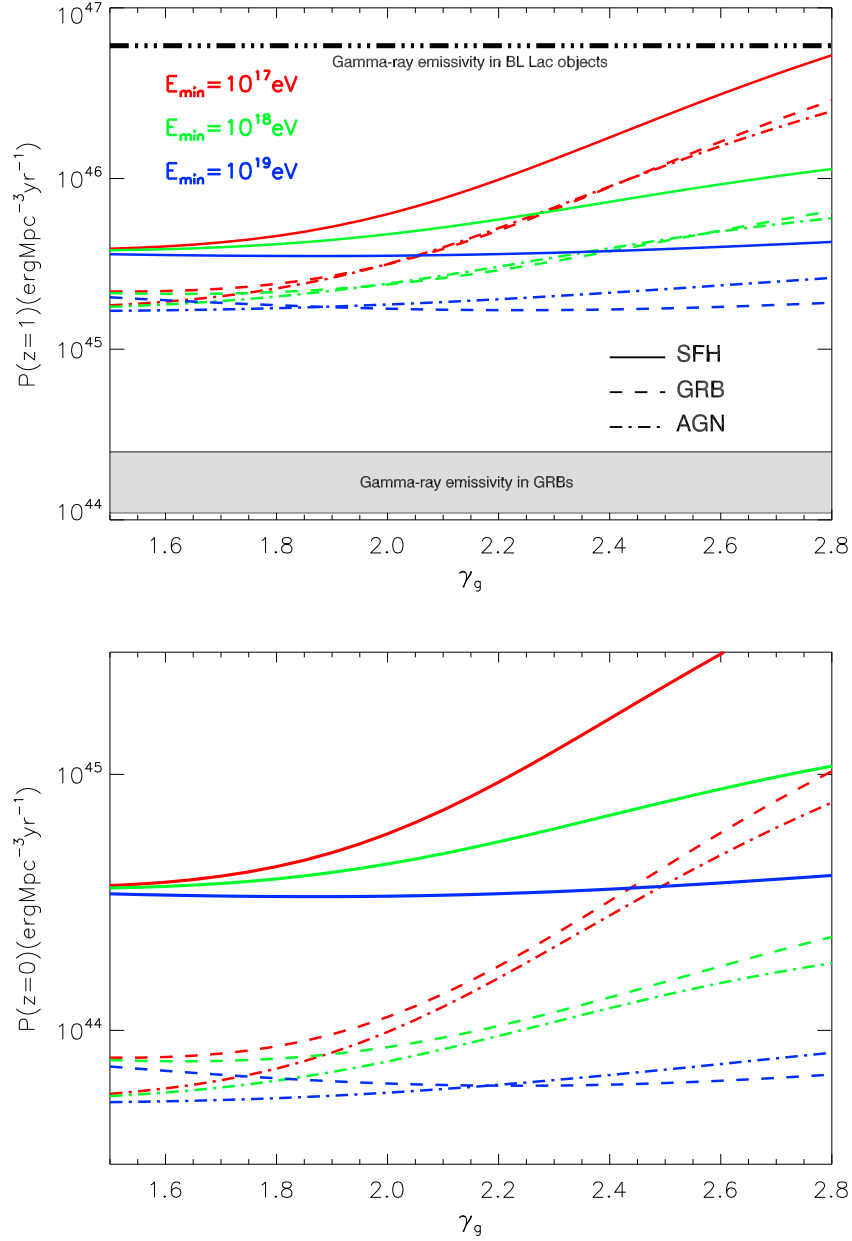,width=12cm} \caption{Upper limits
on the UHECR
 emissivity as a function of the spectral index of the cosmic
ray proton spectrum for different source density evolution
scenarios. The upper panel shows the upper limits on the UHECR
emissivity at  $z=1$,  and the bottom panel shows the  upper limits
on the UHECR emissivity at the present time ($z\ll1$) if the source
density evolution follows $S(z)$ up to the present time. The maximum
energy is fixed at $E_{p,max}=10^{21}{\rm eV}$, while three values
are assumed for the minimum energy. In the upper panel, the
gamma-ray emissivity of candidate UHECR sources (GRBs and BL Lac
objects) at $z=1$ are plotted for comparison. \label{fig2}}
\end{figure}

\begin{figure}
\centering \epsfig{figure=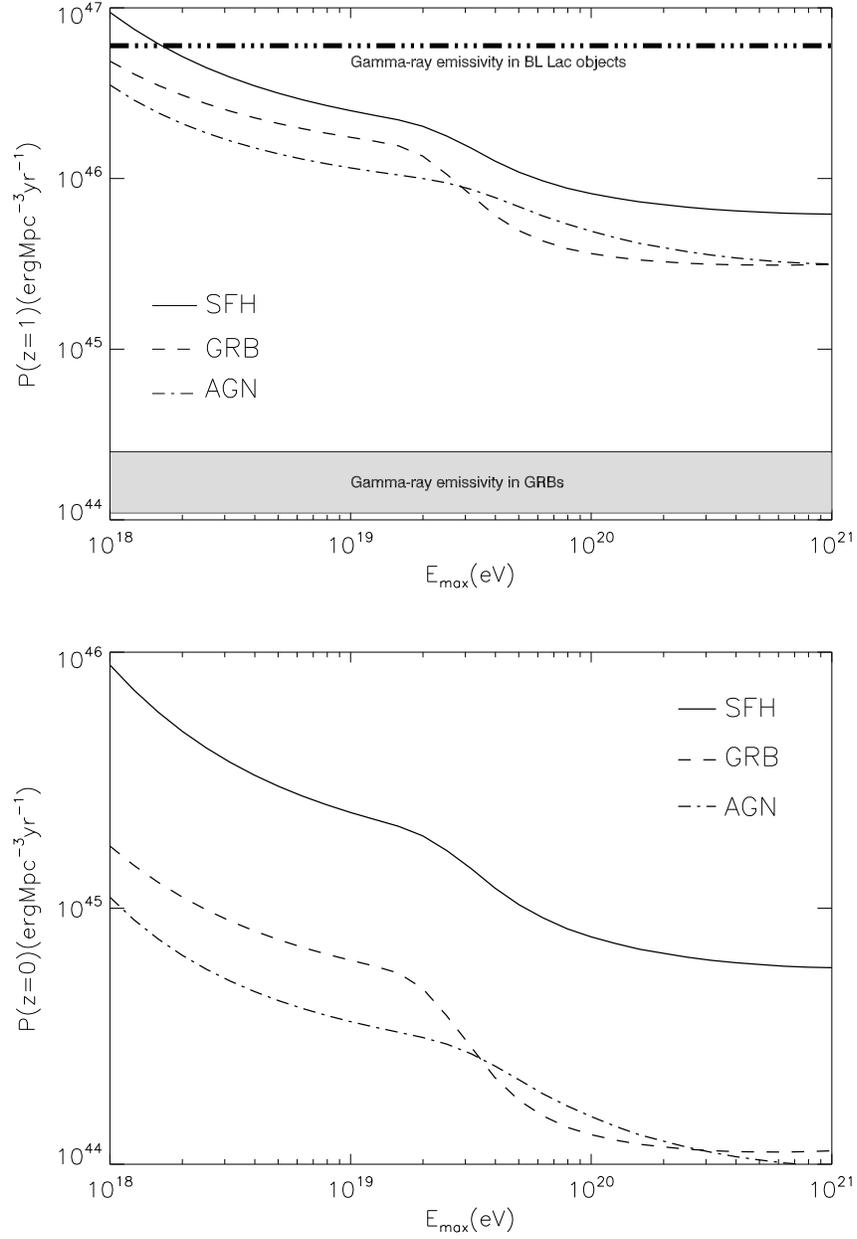,width=12cm} \caption{Upper limits
on the UHECR  emissivity as a function of the maximum energy of the
cosmic ray spectrum for different source density evolution
scenarios.  The spectral index of the the cosmic ray spectrum is
fixed at $\gamma_g=2$ and $E_{p,min}=10^{17}{\rm eV}$.\label{fig3}}
\end{figure}

\begin{figure}
\centering \epsfig{figure=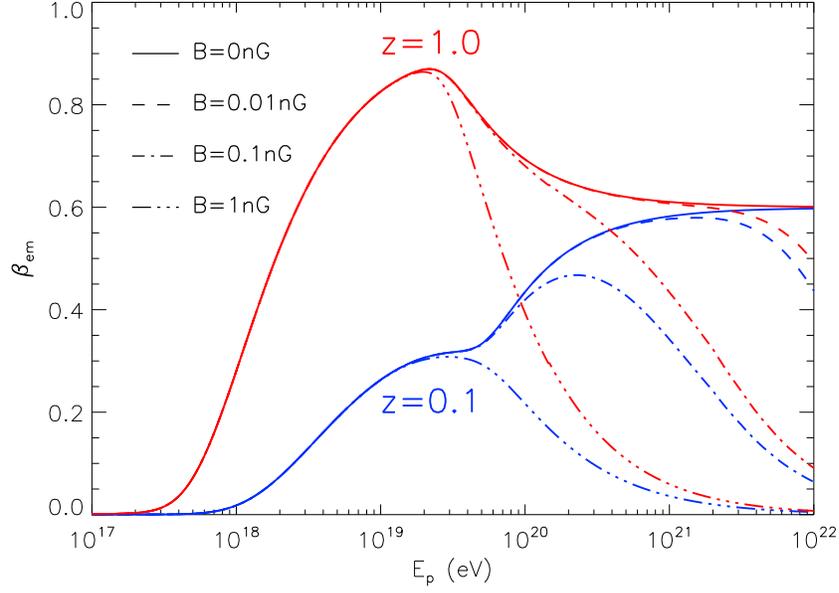,width=12cm} \caption {The
fraction energy loss of UHE protons into the electro-magnetic
component for protons generated at two different redshifts in the
presence of intergalactic magnetic fields. Different lines
correspond to different strength of the intergalactic magnetic
fields. \label{fig4}}
\end{figure}

\begin{figure}
\centering \epsfig{figure=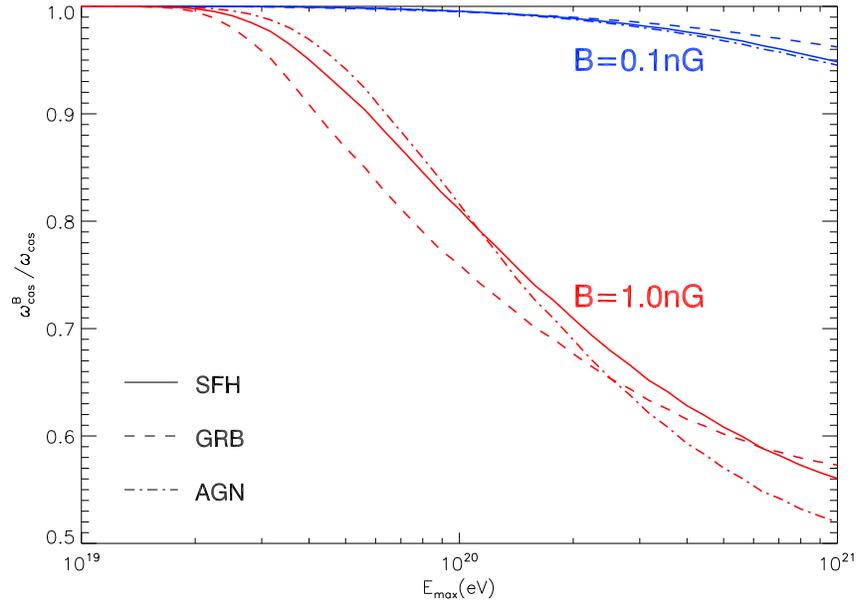,width=12cm} \caption {The ratios
of the cascade energy densities in the presence and in the absence
of intergalactic magnetic fields. The blue and red lines correspond
to the ratios for intergalactic magnetic fields of 0.1 nG and 1 nG,
respectively. $E_{p,min}=10^{17}{\rm eV}$ and $\gamma_g=2$ are used
in the calculation. \label{fig5}}
\end{figure}

\clearpage

\begin{figure}
\centering \epsfig{figure=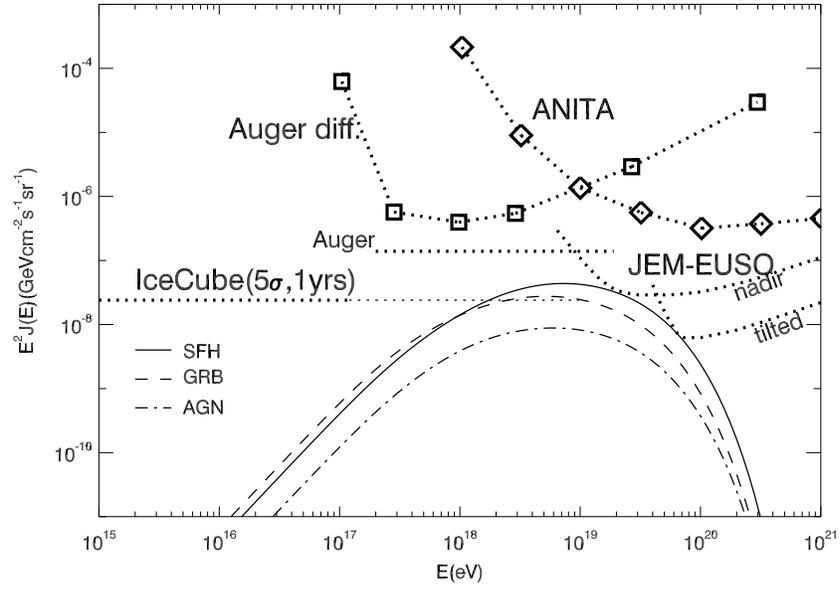,width=12cm}\caption {Upper limits
on the cosmogenic neutrino flux ($\nu_\mu+\bar\nu_\mu$) , derived
from the upper limits of the UHECR emissivity in the universe, for
different source density evolution scenarios. $E_{p,min}=10^{17}{\rm
eV}$, $E_{p,max}=10^{21}{\rm eV}$ and $\gamma_g=2$ are used in the
calculation.\label{fig6}}
\end{figure}

\begin{figure}
\centering \epsfig{figure=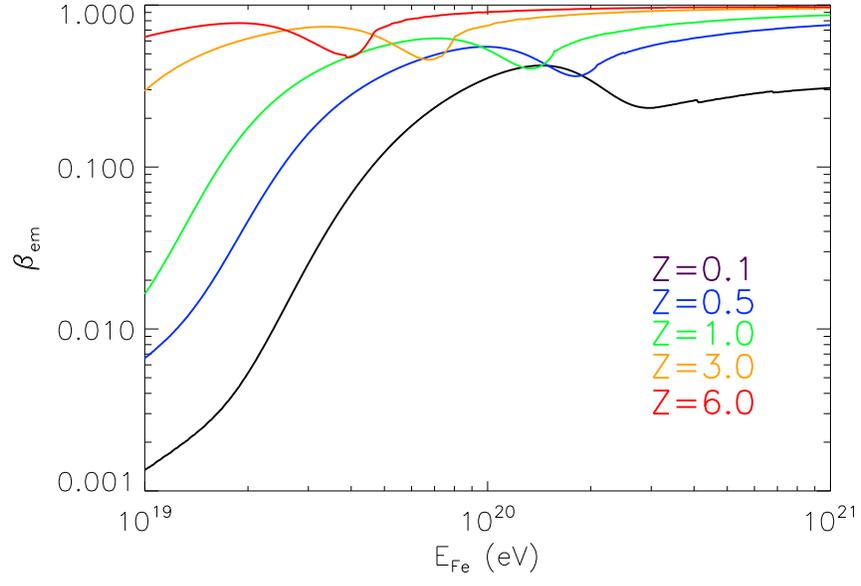,width=12cm} \caption{The fraction
energy loss of UHE iron nuclei  into the electro-magnetic component
for iron nuclei generated at different redshifts.   \label{fig7}}
\end{figure}

\begin{figure}
\centering \epsfig{figure=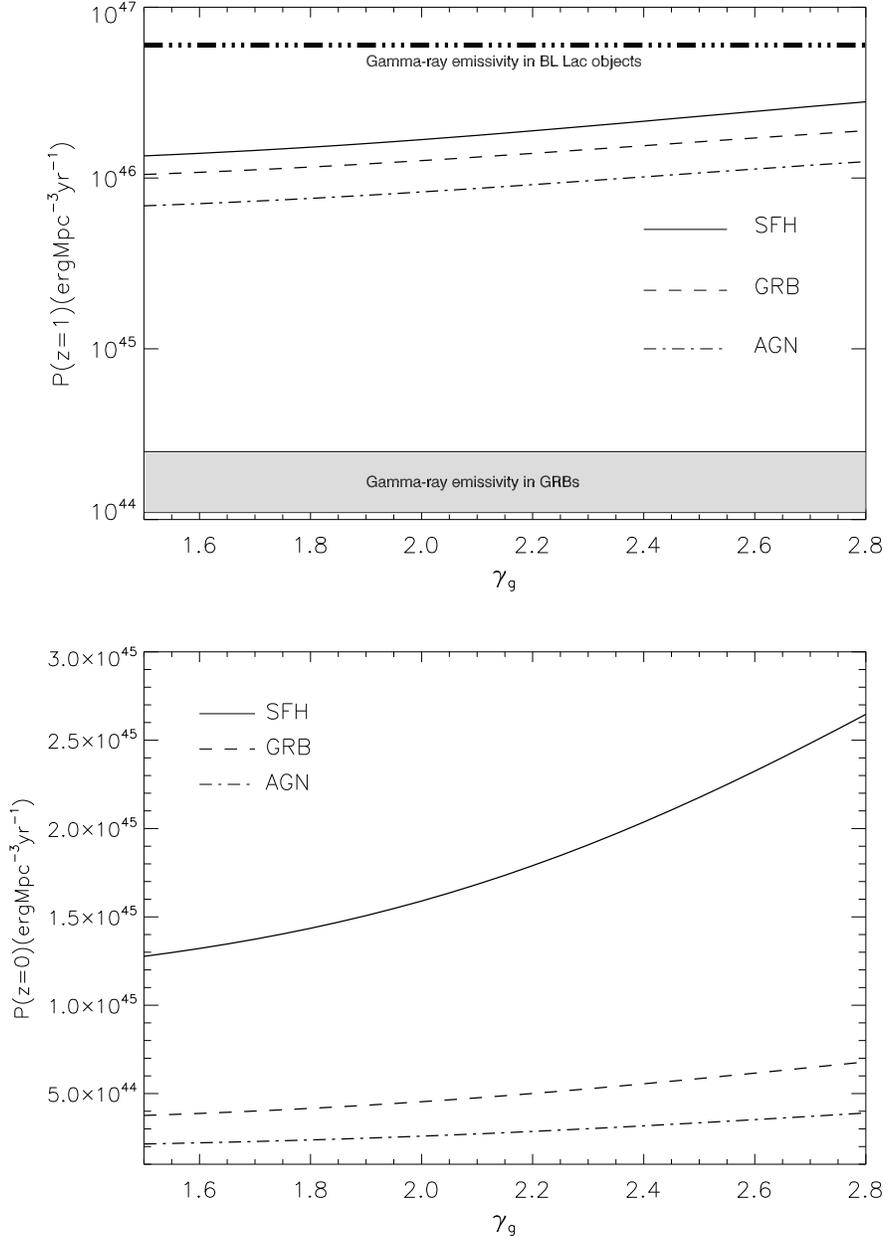,width=12cm} \caption{Upper limits
on the UHECR  emissivity as a function of the spectral index for a
pure iron nuclei composition of UHECRs in the energy range from
$E_{min}=10^{19}$ eV to $E_{max}=10^{21}$ eV. \label{fig8}}
\end{figure}

\end{document}